\documentclass[a4paper,nobibnotes,nofootinbib]{revtex4}
\usepackage{amssymb}
\usepackage{amsmath}
\usepackage{epsfig}
\newcommand{\be}{\begin{equation}}
\newcommand{\ee}{\end{equation}}
\newcommand{\bea}{\begin{eqnarray}}
\newcommand{\eea}{\end{eqnarray}}

\newcommand{\half}{\textstyle{\frac{1}{2}}}
\newcommand{\g}{\gamma}
\newcommand{\f}{\frac}

\newcommand{\bra}{\langle}
\newcommand{\ket}{\rangle}
\newcommand{\pom}{{\mathbb P}}
\newcommand{\qbar}{{\bar{q}}}
\newcommand{\xp}{x_{\mathbb P}}

\newcommand\lr[1]{{\left({#1}\right)}}
\begin{document}

\title{Testing saturation with diffractive jet production in deep inelastic 
scattering}

\author{K. Golec-Biernat}\email{golec@ifj.edu.pl}
\affiliation{Institute of Nuclear Physics, Radzikowskiego 152,
31-342 Krak{\'o}w, Poland\\
Institute of Physics, University of Rzesz\'ow, Rzesz\'ow, Poland}

\author{C. Marquet}\email{marquet@spht.saclay.cea.fr}
\affiliation{Service de Physique Th{\'e}orique, CEA/Saclay,
91191 Gif-sur-Yvette Cedex, France\footnote{%
URA 2306, unit{\'e} de recherche associ{\'e}e au CNRS.}}
\begin{abstract}

We analyse the dissociation of a photon in diffractive deep inelastic scattering
in the kinematic regime where the diffractive mass is much bigger than the 
photon virtuality.
We consider the dominant $q\bar{q}g$ component keeping track of the transverse 
momentum of the
gluon which can be measured as a final-state jet. We show that the diffractive 
gluon-jet production cross-section is strongly sensitive to unitarity 
constraints. In particular, in a model with parton saturation, this 
cross-section is sensitive to the scale at which unitarity effects become 
important, the saturation scale. We argue that the measurement of diffractive 
jets at HERA in the limit of high diffractive mass can provide useful 
information on the saturation regime of QCD.

\end{abstract}
\maketitle

\section{Introduction}
\label{sec:1}

The understanding of diffractive interactions in electron-proton deep inelastic 
scattering (DIS)
has been a great theoretical challenge since diffractive events were observed at 
HERA \cite{h1zeusdata}.
There exist many attempts to describe the diffractive part of the deep inelastic
cross-section within perturbative QCD (for an excellent review, see 
Ref.~\cite{heb}).
One of the most successful approaches is based on the dipole
picture of DIS \cite{dipole,mueller} which expresses the scattering of the 
photon of virtuality
$Q^2$ through its fluctuation into a color singlet $q\bar q$ pair (dipole) of a 
transverse size
$r\!\sim\!1/Q$. That naturally incorporates the description of both inclusive 
and diffractive events into a common theoretical framework 
\cite{nikzak,biapesroy}, as the same dipole scattering amplitudes enter in the 
formulation of the inclusive and diffractive cross-sections.

The dipole approach revealed that the total diffractive cross-section is much 
more sensitive
to large-size dipoles than the inclusive one \cite{golec}. More precisely, it 
showed that
unitarity, and the way it is realized, should be important ingredients of the 
description of
diffractive cross-sections, making those ideal places
to look for saturation effects at small-$x$. The saturation parametrization of 
the dipole scattering amplitude
\cite{golec} was quite successful in describing both the inclusive and 
diffractive structure functions.
In other studies of saturation effects in diffractive DIS, nonlinear evolution
equations for the structure function
have been derived \cite{kovlev,kovwie}, new measurements
proposed \cite{match}, and fits of differents sets of data performed
\cite{munsho,forsh}.

In this paper, we analyse hard diffraction when the proton stays intact after 
the collision and
the mass $M_X$ of the diffractive final state is much bigger
than $Q^2$.  This process is called diffractive photon dissociation. We extend 
the study of \cite{munsho}
by keeping track of the transverse momentum of the final state partons. We 
propose the
measurement of the final state configuration ${\rm X+jet+gap+p}$ in virtual 
photon-proton collisions. In order to connect the measured jet with the 
final-state gluon in our calculations, the jet should form the edge of the 
rapidity gap.
The transverse momentum of the jet provides a hard scale necessary for the use 
of perturbative QCD, making our calculations valid even at very low values of  
$Q^2$.

We express the diffractive cross-section in terms of dipole scattering 
amplitudes, using the results derived
in \cite{cyrille} in the eikonal approximation, valid at very high $\gamma^*p$ 
center-of-mass energy. We
show that in the context of saturation theory,
the transverse momentum distribution of the measured jet is resonant with the 
scale at which the
contributions of large-size dipoles start to be suppressed, called the 
saturation scale.
Using the
parametrization \cite{golec} of saturation effects, we make predictions for the 
kinematic domain of HERA and exhibit
the potential of the diffractive jet production measurement for extracting the 
saturation scale.

The plan of the paper is as follows. In Section II we recall the derivation of 
\cite{cyrille} for
the diffractive production of a gluon off a $q\bar{q}$ dipole.
In Section III, we derive the cross-section for the diffractive photon 
dissociation with  production of a gluon jet and study its model-independent 
properties.
In section IV, we present the saturation model  that we use for the calculation 
of  the jet production cross-section. Section V displays our predictions for the 
HERA energy range, and Section VI contains conclusions.

\section{Diffractive gluon production off a $Q\bar Q$ dipole}
\label{sec:2}

\begin{figure}[ht]
\begin{center}
\epsfig{file=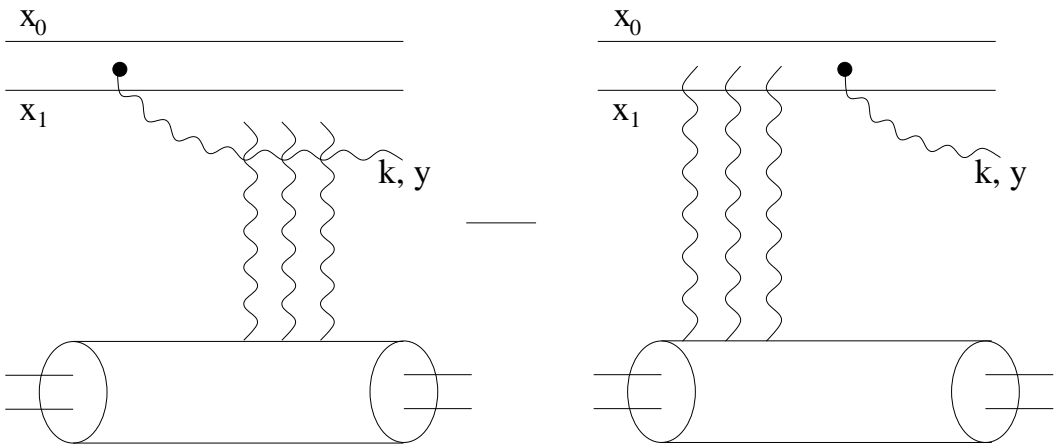,width=14cm}
\caption{Diffractive gluon production off a $q\bar q$ dipole.
$x_0$ and $x_1$ are the transverse coordinates of the quark and antiquark
while $k$ and $y$ are the transverse momentum and rapidity of the
measured gluon. Black points represent emission of a gluon from the dipole and
the vertical wavy lines correspond to a color singlet interaction with a target.
The amplitude for the gluon emission after the interaction comes with a minus 
sign.}
\end{center}
\label{F1}
\end{figure}

In this section we recall the derivation of \cite{cyrille} of the cross section 
for the 
diffractive production of a gluon in the high-energy
scattering of a $q\bar q$ dipole off an arbitrary target. We shall use the 
light-cone
coordinates with the incoming dipole being a right mover and work in the 
light-cone gauge $A_+=0.$ In such a case, when the dipole passes through the
target and interacts with its gauge fields, the dominant couplings are eikonal.
The partonic components of the dipole have frozen transverse coordinates, 
and the gluon fields of the target do not
vary during the interaction. This is justified since the incident dipole 
propagates at nearly the speed of light and its time of propagation through the 
target is shorter than the natural time scale on which the target fields vary.
The effect of the interaction with the target is that the components of the
dipole wavefunction pick up eikonal phases.

In Fig.~1 we present the production of a gluon of
transverse momentum $k$ and rapidity $y$ off a quark-antiquark dipole with 
transverse
coordinates $x_0$ and $x_1$. The transverse size of the
dipole $|x_0\!-\!x_1|$ is supposed to be small in order to justify the use of 
perturbative QCD ($|x_0\!-\!x_1|\!\ll\! 1/\Lambda_{QCD}$). We work in a frame
in which the dipole rapidity is not too large so that the radiation of extra
softer gluons is described by quantum evolution of the target.

The incident hadronic state is a colorless $q\qbar$ dipole state which has the 
following
decomposition on the Fock states:
\be
|d\,\ket\,=\,|d\,\ket_0\,+\,|d g\ket_0
\label{decomp}
\ee
where the bare dipole $|d\ket_0$ is characterised by the wavefunction
\be
|d\,\ket_0\,=\,\sum_{\alpha\bar\alpha}
\f{\delta_{\alpha,\bar{\alpha}}}{\sqrt{N_c}}\,
|\,(\alpha,x_0),(\bar{\alpha},x_1)\,\ket
\ee
with $\alpha$ and $\bar\alpha$ denoting colors of the quark and antiquark,
and $x_0$ and $x_1$ being  their
transverse positions.
The $q\bar qg$ part of the dressed dipole $|dg\ket_0$ is characterised by the 
wavefunction
\be
|dg\ket_0\,=\,\f1{\sqrt{N_c}}\sum_{\alpha\bar\alpha\lambda a}\int d^2z \,dy\,
\f{ig_s}{\pi}\left[\f{\epsilon_\lambda.(x_0\!-\!z)}{|x_0\!-\!z|^2}
-\f{\epsilon_\lambda.(x_1\!-\!z)}{|x_1\!-\!z|^2}
\right]
T_{\bar\alpha \alpha}^a\,
|\,(\alpha,x_0),(\bar\alpha,x_1),(a,\lambda,z,y)\,\ket\ ,\label{softg}\ee
where $(a,\lambda,z,y)$ characterize gluon color, polarization, transverse 
coordinate
and rapidity, respectively. In addition, $\epsilon_\lambda$ is the transverse 
component of the 
gluon polarization vector, and $T^a$ is a generator of $SU(N_c)$
in the fundamental representation. The term in brackets in
(\ref{softg}) is the well-known wavefunction for the emission of a gluon off a
$q\bar q$ dipole \cite{mueller}.  The two contributions correspond to
emission from  the quark and antiquark. The only assumption made to
write down (\ref{softg}) is that the gluon is soft, 
that is its longitudinal momentum fraction with respect to the incident
dipole is small. As already mentioned, we work in the frame in which only bare 
or
one-gluon components need to be considered in the wavefunction $|d\,\ket$.  
Softer
gluons will be included through quantum evolution of the target.

Let us denote by $|t\,\ket$ the initial state of the target. The outgoing state 
is
obtained from the incoming state $|d\,\ket\otimes |t\,\ket$ by the action of the
${\cal S}-$matrix. In the eikonal approximation, ${\cal S}$ acts on quarks and 
gluons as (see for example \cite{coursmuel,kovwie,heb}):
\be
{\cal S}\,(|(\alpha,x)\ket\!\otimes\! |t\ket)\,=\,
\sum_{\alpha'}\left[W_F(x)\right]_{\alpha\alpha'}
|(\alpha',x)\ket\!\otimes\!|t\ket\ ,
\hspace{1.0cm}
{\cal 
S}\,(|(a,\lambda,z,y)\ket\!\otimes\!|t\ket)\,=\,\sum_{b}\left[W_A(z)\right]^{ab}
|(b,\lambda,z,y)\ket\!\otimes\!|t\ket\
\ee
where phase shifts due to the interaction are described by
the eikonal Wilson lines $W_F$ and $W_A$  in the fundamental and adjoint 
representations,
respectively, corresponding to the propagating quarks and gluons. They are given 
by the following
path ordered exponential
\begin{equation}
W_{F,A}(x)\,=\,{\cal P}\exp\left\{ig_s\!\int dz_+\,{\cal 
A}_-^a(x,z_+)\,T_{F,A}^a\right\}
\end{equation}
with ${\cal A}_-^a$ being the target gauge field, and $T_{F,A}$ are generators
of the color group in the fundamental and adjoint representation.
Thus, the state
$|\Psi_{out}\ket = {\cal S}\,|d\ket\! \otimes\! |t\ket$, emerging from the 
eikonal
interaction, is given by  
\be
|\Psi_{out}\ket = |\Psi_1\ket+|\Psi_2\ket
\ee
 with
\bea
\label{psi1}
|\Psi_1\ket
\!\!&=&\!\!
\f1{\sqrt{N_c}}\sum_{\alpha\bar\alpha}
\left[W_F^\dagger(x_1)\,W_F(x_0)\right]_{\bar{\alpha}\alpha}
|(\alpha,x_0),(\bar{\alpha},x_1)\ket\otimes|t\ket\ ,
\\\nonumber
\\\nonumber
\label{psi2}
|\Psi_2\ket
\!\!&=&\!\!
\f1{\sqrt{N_c}}\sum_{\alpha\bar\alpha\lambda b}\int d^2z\, dy\
\f{ig_s}{\pi}
\left[\f{\epsilon_\lambda\cdot(x_0\!-\!z)}{|x_0\!-\!z|^2}
-\f{\epsilon_\lambda\cdot(x_1\!-\!z)}{|x_1\!-\!z|^2}\right]
\\\nonumber
\\
&&\times
\left[W_F^\dagger(x_1)\,T^a\,W_F(x_0)\right]_{\bar{\alpha}\alpha}
\left[W_A(z)\right]^{ab}\,
|(\alpha,x_0),(\bar\alpha,x_1),(b,\lambda,z,y)\ket\otimes|t\ket\ .
\eea
This outgoing state would be the one to consider to compute inclusive 
cross-sections, 
as no restrictions on the final state have been imposed. To compute diffractive 
cross-sections, 
one has to project the outgoing state $|\Psi_{out}\ket$ on the subspace of 
color-singlet
states. We have defined diffractive processes as ones in which the target does 
not break up,
therefore one also has to project the outgoing state on the subspace spanned by 
the target 
state $|t\ket$. Those projections are described in detail in \cite{cyrille}. 
They create the rapidity gap, preventing the emissions of gluons softer than the 
one described by (\ref{psi2}).
Let us denote the resulting state $|\Psi_{diff}\ket,$ it is given by:
\be
|\Psi_{diff}\ket = |\Psi_1^d\ket+|\Psi_2^d\ket
\ee
with
\bea
|\Psi_1^d\ket
\!\!&=&\!\!
\f1{N_c}\bra t|\mbox{Tr}
\lr{W_F^\dagger(x_1)W_F(x_0)}|t\ket\,|d\ket_0\otimes|t\ket\ ,
\\\nonumber
\\\nonumber
|\Psi_2^d\ket
\!\!&=&\!\!
\f1{C_FN_c}\f1{\sqrt{N_c}}\sum_{\alpha\bar\alpha\lambda
a}\int d^2z dy\ \f{ig_s}{\pi}
\left[\f{\epsilon_\lambda\cdot(x_0\!-\!z)}{|x_0\!-\!z|^2}
-\f{\epsilon_\lambda\cdot(x_1\!-\!z)}{|x_1\!-\!z|^2}\right]
\\\nonumber
\\
&&\times\,\bra t|\mbox{Tr}\lr{W_F^\dagger(x_1)\,T^b\,W_F(x_0)\,T^c}
\left[W_A(z)\right]^{bc}|t\ket\,T^a_{\bar\alpha\alpha}\,
|(\alpha,x_0),(\bar\alpha,x_1),(a,\lambda,z,y)\ket\otimes|t\ket\ .
\eea

The state $|\Psi_2^d\ket$ represents the first contribution 
pictured in Fig.1 when the interaction happens after the emission of the gluon. 
The second contribution when the interaction happens before the gluon emission
is part of $|\Psi_1^d\ket$.  In order to see that, one has to substitute
$|d\ket_0\!=\!|d\ket\!-\!|dg\ket_0$ in $|\Psi_1^d\ket:$ the term that comes 
with 
$|d\ket$ 
is the contribution of elastic scattering while the term that comes with 
$|dg\ket_0$ and a 
minus sign represents the second contribution of Fig.1. One can drop the elastic 
part since 
it does not contribute to gluon production and write:
\bea
|\Psi_{diff}\ket
\!\!&=&\!\!
\f1{C_FN_c\sqrt{N_c}}\sum_{\alpha\bar\alpha\lambda a}\int d^2z\, dy\
\f{ig_s}{\pi}
\left[\f{\epsilon_\lambda\cdot(x_0\!-\!z)}{|x_0\!-\!z|^2}
-\f{\epsilon_\lambda\cdot(x_1\!-\!z)}{|x_1\!-\!z|^2}\right]
\Phi(z)\,T^a_{\bar\alpha \alpha}\,
|(\alpha,x_0),(\bar\alpha,x_1),(a,\lambda,z,y)\ket\otimes|t\ket
\eea
with
\be
\Phi(z)=\bra t|\mbox{Tr}\lr{W_F^\dagger(x_1)T^aW_F(x_0)T^b}
W_A^{ab}(z)|t\ket
-C_F\bra t|\mbox{Tr}\lr{W_F^\dagger(x_1)W_F(x_0)}|t\ket\ .\label{phi}
\ee
From this final state, one calculates the diffractive cross-section for the 
production 
of a gluon of transverse momentum $k$ and rapidity $y$ using
the following formula:
\be
\f{d\sigma_{diff}}{d^2kdy}(x_{01})=\f1{2(2\pi)^3}\int d^2b
\sum_{\lambda=\pm}\sum_{c=1}^{N_c^2-1}\,
\bra\Psi_{diff}|a_{c,\lambda}^\dagger(k,y)\,a_{c,\lambda}(k,y)|\Psi_{diff}\ket
\label{csec}
\ee
where $a_{c,\lambda}^\dagger(k,y)$ and $a_{c,\lambda}(k,y)$ are respectively
the creation and annihilation operators of a gluon with color $c$, polarization
$\lambda$, rapidity $y$ and transverse momentum $k$. The quantity 
$x_{01}\!=\!x_0\!-\!x_1$
is the transverse size of the incoming dipole, and $b\!=\!(x_0\!+\!x_1)/2$ is 
the
impact parameter. The cross-section is computed in details in \cite{cyrille}, 
the final result is:
\be
\f{d\sigma^{diff}}{d^2kdy}(x_{01})=\f{\alpha_s}{\pi^2C_FN_c^2}
\int d^2b\int\f{d^2z_1}{2\pi}\f{d^2z_2}{2\pi}\ 
e^{ik\cdot (z_2\!-\!z_1)}
\left[\f{x_0\!-\!z_1}{|x_0\!-\!z_1|^2}-\f{x_1\!-\!z_1}{|x_1\!-\!z_1|^2}\right].
\left[\f{x_0\!-\!z_2}{|x_0\!-\!z_2|^2}-\f{x_1\!-\!z_2}{|x_1\!-\!z_2|^2}\right]
\Phi(z_1)\Phi^*(z_2)
\label{finaldiff}\ee
with $\Phi$ given by formula (\ref{phi}). Making use of the following identity
\be
2\mbox{Tr}\lr{W_F^\dagger(x_1)\,T^a\,W_F(x_0)T^b}
\left[W_A(z)\right]^{ab}=\mbox{Tr}\lr{W_F^\dagger(x_1)W_F(z)}
\mbox{Tr}\lr{W_F^\dagger(z)W_F(x_0)}
-\f1{N_c}\mbox{Tr}\lr{W_F^\dagger(x_1)W_F(x_0)}\ ,\ee
one is able to rewrite eq.~(\ref{phi}) in terms of the following 
$\cal{S}-$matrices:
\be
S(x_0,x_1)=\f1{N_c}\bra t|\mbox{Tr}\lr{W_F^\dagger(x_1)W_F(x_0)}|t\ket
\ee
for the scattering of a dipole with the quark and antiquark at transverse 
coordinates $x_0$ and
$x_1$ respectively, and
\be
S^{(2)}(x_0,z,x_1)=\f1{N_c^2}\bra t|\mbox{Tr}\lr{W_F^\dagger(x_1)W_F(z)}
\mbox{Tr}\lr{W_F^\dagger(z)W_F(x_0)}|t\ket
\ee
for the scattering of two dipoles, one with the quark and antiquark at 
transverse coordinates 
$x_0$ and $z$ and the other with the quark and antiquark at transverse 
coordinates $z$ and
$x_1$ . One obtains:
\be
\f2{N_c^2}\Phi(z)=S^{(2)}(x_0,z,x_1)-S(x_0,x_1)\ .\label{smat}
\ee
We have not specified the rapidity dependence of the $\cal{S}-$matrices, it is 
the
rapidity at which the target is evolved. If $Y$ is the total rapidity, then the 
$\cal{S}-$matrices in $\Phi$ depend on $Y\!-\!y.$

Note that if one considers that the target is a nucleus, and that each 
scattering on the 
nucleons happens via a two-gluon exchange, then the target averages in
(\ref{phi}) are computable (see {\it e.g.} \cite{kovwie}) and one recovers the
result of \cite{kovch}. Formulae (\ref{finaldiff}) and (\ref{phi}) are a 
generalization to any target that includes all numbers of gluon exchanges.
Note also that if one writes eq.~(\ref{smat}) in terms of $T-$matrices 
(${\cal{S}}=1-T$),
one recovers the two-gluon exchange approximation calculated in 
\cite{bart,kopdiff} 
by neglecting the term proportional to $T^2$.
Let us now apply formulae (\ref{finaldiff}) and (\ref{phi}) to diffractive 
photon dissociation.

\section{Diffractive photon dissociation}
\label{sec:3}

In deep inelastic scattering, a photon of virtuality $Q^2$ collides
with a proton. In an appropriate frame, called the dipole frame, the virtual 
photon undergoes the hadronic interaction via a fluctuation into a dipole. The  
wavefunctions $|\psi^\gamma_T|^2$ and $|\psi^\gamma_L|^2$, describing the 
splitting of the photon 
on the dipole, are given by
\be
\begin{split}
|\psi^{\g}_T(r,\alpha;Q)|^2&=
\frac{\alpha_{em}N_c}{2\pi^2}\sum_f e_f^2
\left((\alpha^2+(1-\alpha)^2)\varepsilon_f^2K_1^2(\varepsilon_f |r|)
+m_f^2 K_0^2(\varepsilon_f |r|)\right)\\
|\psi^{\gamma}_L(r,\alpha;Q)|^2&=
\frac{\alpha_{em}N_c}{2\pi^2}\sum_f e_f^2
4Q^2 \alpha^2(1-\alpha)^2 K_0^2(\varepsilon_f|r|)
\end{split}\label{phwf}
\ee
for a transversely and longitudinally polarized photon, respectively. In the 
above 
$\varepsilon_f=\sqrt{\alpha(1\!-\!\alpha)Q^2\!+\!m_f^2}$ with 
$m_f$ the mass of the quark $f,$ $r$ is the transverse 
size of the $q\bar q$ pair and $\alpha$ (resp. $1\!-\!\alpha$) is the 
longitudinal momentum fraction of the antiquark (resp. quark). The dipole then 
interacts with the target 
proton and one has the following factorization
\be
\sigma^{\g^*p}=\int d^2 r 
\int_0^1 d\alpha
\left(|\psi_T^{\g}(r,\alpha;Q)|^2
+|\psi_L^{\g}(r,\alpha;Q)|^2
\right) \sigma(r,\alpha)\label{factor}
\ee
which relates a cross-section for an incident photon $\sigma^{\g^*p}$ to the 
corresponding 
cross-section with an incident dipole $\sigma(r,\alpha).$ 
In the leading logarithmic approximation we are interested in, the dipole 
cross-sections
do not depend on $\alpha$ and one defines then:
\be
\label{eq:phi}
\phi(r,Q)\,=\int 
d\alpha\left(\,|\psi_T^{\g}(r,\alpha;Q)|^2+|\psi_L^{\g}(r,\alpha;Q)|^2\right).
\ee
We are going to use the factorization formula (\ref{factor}) to compute the 
diffractive photon
dissociation cross-section.
\begin{figure}[ht]
\begin{center}
\epsfig{file=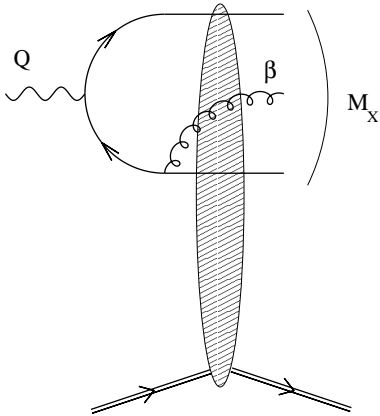,width=5cm}
\caption{Diffractive photon dissociation in virtual photon-proton collisions. 
This is the dominant
contribution to the diffractive cross-section when the final-state diffractive 
mass $M_X$ 
is much bigger than the photon virtuality $Q.$}
\end{center}
\label{F2}
\end{figure}

In diffractive deep inelastic scattering, the proton gets out of the collision 
intact and there is a rapidity gap between that proton and the final state $X,$ 
see Fig.~2.
If the final state diffractive mass $M_X$ is much bigger than 
$Q,$ then the dominant contributions to the final state come from the $q\bar q 
g$ component of the photon wavefunction or from higher Fock states, 
i.e. from the photon dissociation. 
By contrast, if $M_X\!\ll\!Q,$ the dominant contribution 
comes from the $q\bar q$ component. In this paper we investigate 
the $q\bar q g$ component, in the kinematical region where 
\be
\beta\equiv\f{Q^2}{Q^2+M_X^2}\ll1\ .
\ee
One can easily express the diffractive mass of the $q\bar qg$ final state
in terms of the kinematical variables  of the partons:
\begin{equation}
M_X^2=\frac{k^2_q}{z_q}+\frac{k^2_{\bar q}}{z_{\bar q}}+\frac{k_g^2}{z_g}
\end{equation}
where $z_q$, $z_{\bar q},$ and $z_g$ are the  longitudinal momentum  fractions 
of the quark, antiquark and gluon, respectively, ($z_q\!+\!z_{\bar 
q}\!+\!z_g\!=\!1$), and $k_q$, $k_{\bar q},$ and $k_g$ are their transverse 
momenta ($k_q\!+\!k_{\bar q}\!+\!k_g\!=\!0$). Several kinematical configurations 
can provide $\beta\!\ll\!1,$ however 
the configuration that gives the dominant contribution to the cross-section is:
\begin{equation}
\frac{k_g^2}{z_g}\gg\frac{k^2_q}{z_q},\ \frac{k^2_{\bar q}}{z_{\bar q}},\ Q^2\ .
\label{approx}\end{equation}
This is due to the infrared singularity of QCD. Indeed, as we shall see later, 
this ordering corresponds to the resummation of Feynman diagrams in the leading 
logarithmic $\log(1/\beta)$ approximation. It also corresponds to a final-state 
configuration
where the gluon jet is the closest to the gap.

In the previous section, we obtained the diffractive cross-section for the 
production of a 
gluon with transverse momentum $k$ and rapidity $y$ in the collision of a dipole 
of tranverse 
size $x_{01}$ with the target proton in the 
approximation (\ref{approx}). The result reads
\be
\label{eq:sigd}
\f{d\sigma^{diff}}{d^2kdy}(x_{01})\,=\,\f{\alpha_sN_c^2}{4\pi^2C_F}
\int d^2b\ A(k,x_0,x_1;\Delta\eta)\cdot A^*(k,x_0,x_1;\Delta\eta)
\ee
where $x_0=b+x_{01}/2$, $x_1=b-x_{01}/2$, $Y$ is   the total rapidity and
$\Delta\eta=Y-y$ is the rapidity gap. The two-dimensional vector $A$ is given by
\be
A(k,x_0,x_1;\Delta\eta)=\int\f{d^2z}{2\pi}\ e^{-ik\cdot z}
\left[\f{z\!-\!x_0}{|z\!-\!x_0|^2}-\f{z\!-\!x_1}{|z\!-\!x_1|^2}\right]
\lr{S^{(2)}(x_0,z,x_1;\Delta\eta)-S(x_0,x_1;\Delta\eta)}
\label{ampla}
\ee
where $S(x_0,x_1;\Delta\eta)$ is the elastic ${\cal S}-$matrix for the collision 
of the 
dipole $(x_0,x_1)$ on the target proton evolved at the rapidity $\Delta\eta$, 
and 
$S^{(2)}(x_0,z,x_1;\Delta\eta)$ is the elastic ${\cal S}-$matrix for the 
collision of two dipoles $(x_0,z)$ and $(z,x_1)$. 
These formulae are valid at leading logarithmic accuracy in $y=\log(1/\beta)$ as 
pointed 
out before. Indeed, after squaring the term in brackets
in (\ref{ampla}), one obtains the BFKL kernel. The first (resp. second) term in
the brackets corresponds to the emission of the gluon at transverse position $z$ 
from the quark (resp. antiquark) at transverse position $x_0$ (resp. $x_1$). The 
$S^{(2)}$ (resp. $S$) term represents the case where the interaction with the 
target takes place after (resp. before) the emission of the gluon.  We shall 
refer to it as the real (resp. virtual) term.

Let us introduce the usual kinematics of diffractive DIS: $Y=\log(1/x)$ and
$\Delta\eta=\log(1/x_{\pom})$ with 
\be
x=\f{Q^2}{Q^2+W^2}\ , \hspace{2cm} x_{\pom}=\f{Q^2+M_X^2}{Q^2+W^2}\ ,
\ee
where $W^2$ is the center-of-mass energy of the photon-proton collision.
Using the factorization (\ref{factor}), one obtains the $q\bar qg$ component of 
the 
diffractive cross-section in the virtual photon-proton collision:
\be
\f{d\sigma_{diff}^\g}{d^2k\ dM_X} =\f{2M_X}{M_X^2+Q^2}
\int d^2x_{01}\, \phi(|x_{01}|,Q)\,
\f{d\sigma^{diff}}{d^2kdy}(x_{01})\label{cs}\,.
\ee
with the photon wavefunction given by formula (\ref{eq:phi}) and the dipole 
cross-section given by
formulae (\ref{eq:sigd}) and (\ref{ampla}). It is differential with respect to 
the diffractive 
mass $M_X$ and to the final-state gluon transverse momentum $k,$ which can be 
identified with the
transverse momentum of the jet which is the closest to the rapidity gap. Note 
that the transverse momentum $k$ provides the hard scale, so that we can apply 
our formulae to even low values of $Q^2.$

Let us make some general comments on the $k-$dependence of the cross-section
(\ref{cs}).
\begin{itemize}
\item When $k\to 0$.\\
The amplitude $A$ given by eq.~(\ref{ampla}) takes a constant value. The 
infrared divergences
a priori appearing for the virtual term cancel between the $x_0$ and $x_1$ part 
and the dominant contribution to $A$ is determined by the large $z$ behavior of 
$S^{(2)}(x_0,z,x_1;\Delta\eta)$. In particular, the value of $z$ at which
$S^{(2)}(x_0,z,x_1;\Delta\eta)$ starts decreasing to zero plays the role of a 
natural cutoff and determines the value of $A$. The constant value of the
cross-section (\ref{cs}) at $k=0$ is then very sensitive to the way that
unitarity sets in.
\item When $k\to \infty$.\\ 
The amplitude $A$ decreases as $1/k^2$. By changes of variables, one can
write
\be
A(k,x_0,x_1;\Delta\eta)=\f{e^{-ik.x_0}}{|k|}\int\f{d^2z}{2\pi}\ e^{-ik.z/|k|}
\f{z}{|z|^2}\lr{S^{(2)}(x_0,z/|k|+x_0,x_1;\Delta\eta)-S(x_0,x_1;\Delta\eta)}
-(x_0\leftrightarrow x_1).
\ee
Then taking $k\!\rightarrow\!\infty$, and using
\be
S^{(2)}(x_0,x_0,x_1;\Delta\eta)=S^{(2)}(x_0,x_1,x_1;\Delta\eta)
=S(x_0,x_1;\Delta\eta)\,,
\ee
one sees that the $1/|k|$ term vanishes leaving the dominant contribution 
behaving as $1\!/\!k^2:$
\be
A(k,x_0,x_1;\Delta\eta)=\f1{k^2}
\lr{e^{-ik.x_0}\left.\nabla_zS^{(2)}\right|_{x_0}
-e^{-ik\cdot x_1}\left.\nabla_zS^{(2)}\right|_{x_1}}\ .
\ee
Squaring and integrating the impact parameter one obtains
\be
\f{d\sigma^{diff}}{d^2k\ dy}(x_{01})\,\propto\,
\frac{F(|x_{01}|)+G(|x_{01}|)\cos(k\cdot x_{01})}{k^4}
\ee
with $F$ and $G$ depending on the precise form of $S^{(2)}.$ When integrating 
over the angle of $x_{01}$ in (\ref{cs}), the $G$ part becomes suppressed due to
the $J_0(k|x_{01}|)$ function, and the cross-section then falls as $1/k^4$.
\end{itemize}

These features are general, independent of the form of the ${\cal S}-$matrices.
If one looks at the behavior of the observable
\be
k^2 \f{d\sigma_{diff}^\g}{d^2k\ dM_X}
\label{obs}
\ee
as a function of the gluon transverse momentum $k$,
it is going to rise as $k^2$ for small values of $k$ and
fall as $1/k^2$ for large values of $k$. A maximum will occur for a value
$k_0$ which is related to the inverse of the typical size for which the ${\cal
S}-$matrices approach zero; in other words, the maximum $k_0$ will reflect the 
scale
at which unitarity sets in. We want to explore this phenomenon in the framework
of the theory of parton saturation where unitarity is realized perturbatively.

\section{Saturation model for the ${\cal S}-$matrices}
\label{sec:4}

The exact form of the ${\cal S}-$matrices is unknown, and we have to consider 
models
in order to produce values of the observable (\ref{obs}) at any value of $k$.
For this purpose we consider the following model, inspired by the GBW 
parameterization \cite{golec}
of parton saturation effects:
\bea
\label{eq:S1}
S(x_0,x_1;\Delta\eta)\!&=&\!\Theta(R_p\!-\!|b|)\,\,{\rm 
e}^{-Q_s^2(x_{\pom})x_{01}^2/4}
\,+\,\Theta(|b|\!-\!R_p)\ ,
\\\nonumber
\\
\label{eq:S2}
S^{(2)}(x_0,z,x_1;\Delta\eta)\!&=&\!\Theta(R_p\!-\!|b|)\,\,
{\rm e}^{-Q_s^2(x_{\pom})(x_0\!-\!z)^2/4}\,\,
{\rm e}^{-Q_s^2(x_{\pom})(z\!-\!x_1)^2/4}\,+\,\Theta(|b|\!-\!R_p)\ ,
\eea
where $R_p$ is the radius of the proton. $Q_s$ is the saturation scale, the 
basic quantity 
characterizing saturation effects \cite{glr,glr+,mv,jimwlk}. It is a rising 
function of energy 
through its $\xp-$dependence.
The $b-$dependence of the $S-$ matrices is justified if the
dipole sizes $|x_{01}|,$ $|x_0\!-\!z|$ and $|z\!-\!x_1|$ contribute only when 
they are much smaller than $R_p.$ That is we assume that $x_{\pom}$ is always 
such that $Q_s(x_{\pom})\!\gg\!\Lambda_{QCD}$. Note that the model \eqref{eq:S2} 
for $S^{(2)}$ neglects correlations between the two dipoles, as it is a product 
of two $S$'s.

Interestingly enough, since the ${\cal S}-$matrices 
(\ref{eq:S1}) and (\ref{eq:S2}) are Gaussians, one can
analytically compute the amplitude $A$ given by eq.~(\ref{ampla}). The details
of the derivation and the final result (\ref{A9}) are presented in Appendix A.
With these results, we obtain for the product $A\cdot A^*$:
\bea\nonumber
\label{AA}
&& A(k,x_0,x_1;\Delta\eta)\cdot 
A^*(k,x_0,x_1;\Delta\eta)\,=\,\Theta(|b|\!-\!R_p)\,\,
\f{x_{01}^2}{4k^2}\,\, {\rm e}^{-x_{01}^2Q_s^2/2}
\\\nonumber
\\
&&~~~~~~~~~~~~~~~~~~~~~~~~~~~\times\,
\f{\left|2\lr{\cos(k\cdot x_{01}/2)-{\rm e}^{-k^2/(2Q_s^2)+\,Q_s^2x_{01}^2/8}}\ 
k
+\sin(k\cdot x_{01}/2)Q_s^2\ x_{01}\right|^2}
{(k^2/Q_s^2-Q_s^2x_{01}^2/4)^2+(k\cdot x_{01})^2}\,,
\eea
where we suppress the dependence of $Q_s$ on $x_{\pom}$
in the notation. Because of the theta function, the
$b-$integration in eq.~(\ref{eq:sigd}) gives a factor
$\pi R_p^2 \equiv \sigma_0/2$. The 
result of the integration is then a function of $Q_s$ and the two-dimensional 
vectors $k$ and  $x_{01}$. Inserting (\ref{AA}) into (\ref{cs}), one finally 
writes
\bea\nonumber
\label{fincs}
k^2M_X\f{d\sigma_{diff}^\g}{d^2k\ dM_X}\!&=&\!\f{\alpha_s 
N_c^2\sigma_0}{4\pi^2C_F}\,
\f{M_x^2}{M_x^2+Q^2}
\int \f{r\ dr\ d\theta}
{(k/(rQ_s^2) - rQ_s^2/(4k))^2+\cos^2\theta}\,\,\phi(r,Q)\,\,{\rm 
e}^{-r^2Q_s^2/2}
\\\nonumber
\\\nonumber
&\times&\bigg[\lr{\cos\lr{\half{kr} \cos\theta}\,-\,
{\rm e}^{-k^2/(2Q_s^2)+\,Q_s^2r^2/8}}^2
+\f{Q_s^4r^2}{4k^2}\sin^2\lr{\half{kr} \cos\theta}
\\\nonumber
\\
&&~~~+\,\f{rQ_s^2}k \cos\theta\,\sin\lr{\half{kr} \cos\theta}
\lr{\cos\lr{\half{kr} \cos\theta} - {\rm e}^{-k^2/(2Q_s^2)+ Q_s^2r^2/8}}\bigg]
\eea
where now $k=|k|$, and $\phi(r,Q)$ is given by eq.~(\ref{eq:phi}).
The $\alpha$-integration to calculate $\phi(r,Q)$ and the integrations over $r$ 
and
$\theta$ can  easily be done numerically.


\section{Numerical analysis and its implications}
\label{sec:5}

Let us analyse the $k-$dependence of the diffractive cross-section 
(\ref{fincs}).
For this purpose, we define the scaled diffractive cross section
\be
\label{eq:scaled}
\sigma^{scaled}(k,Q^2,Q_s)\,=\,\f{1}{\alpha_s\/\sigma_0}\,
\left(\f{M_X^2+Q^2}{M_X^2}\right)
M_X\,\f{d\sigma_{diff}^\g}{d^2k\ dM_X}\,,
\ee
which allows us to leave aside the problem of uncertainties due to $\alpha_s$ 
and $\sigma_0.$
The inclusion of $\alpha_s$ and $\sigma_0,$ which have constant values in the 
kinematical domain we consider here, in the actual observable (\ref{obs}) of 
course will not change the following discussion. In addition to the gluon 
transverse momentum $k$,
$\sigma^{scaled}$ is a function of two variables: the photon virtuality $Q^2$ 
and the
saturation scale $Q_s$. We have to keep in mind that the diffractive cross
section (\ref{fincs}) was derived under the assumption that $M_X^2\!\gg\! Q^2$,
thus the factor in the brackets on the r.h.s. of eq.~(\ref{eq:scaled}) is close 
to one.

\begin{figure}[ht]
\begin{center}
\epsfig{file=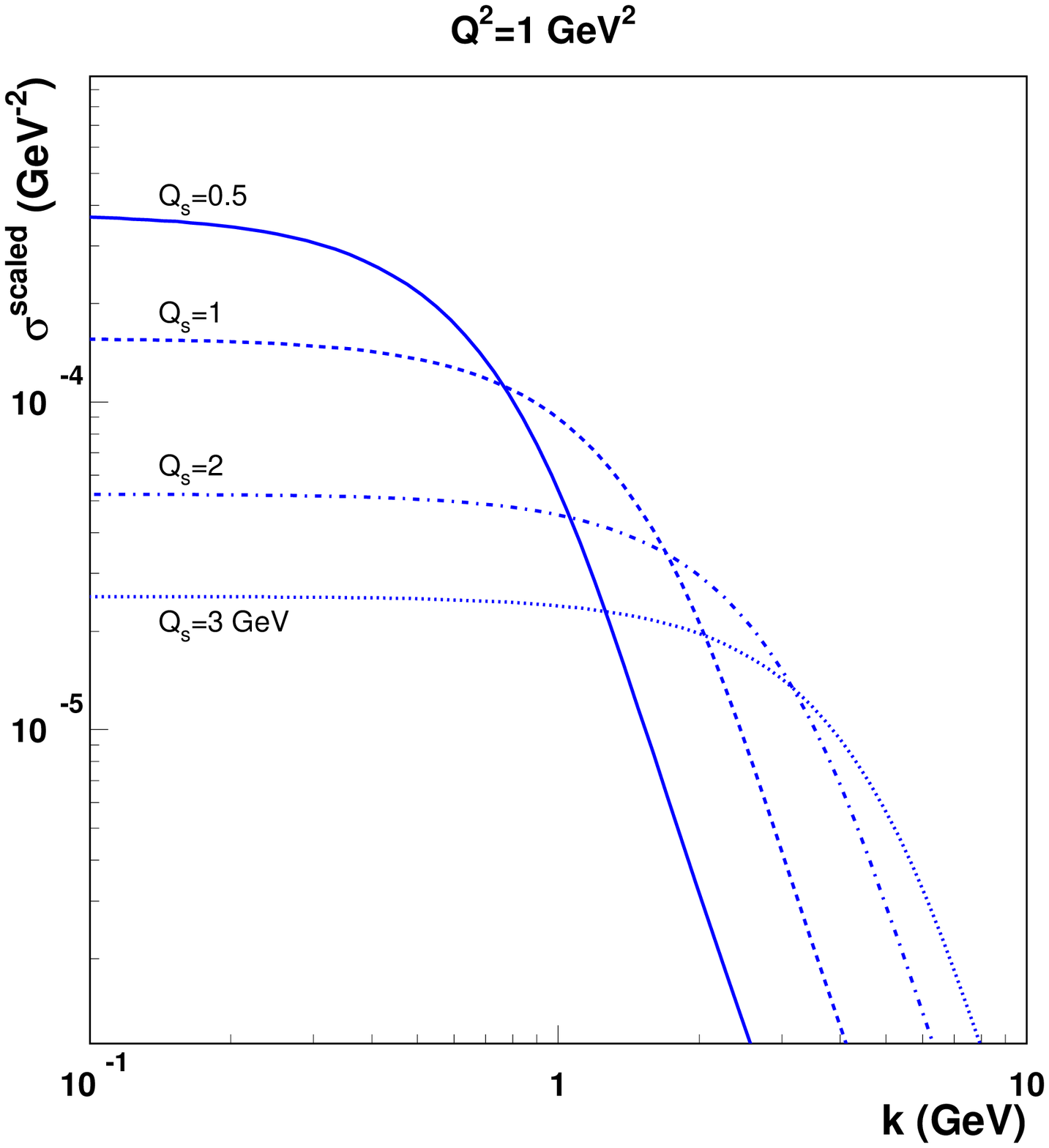,width=8.5cm}
\epsfig{file=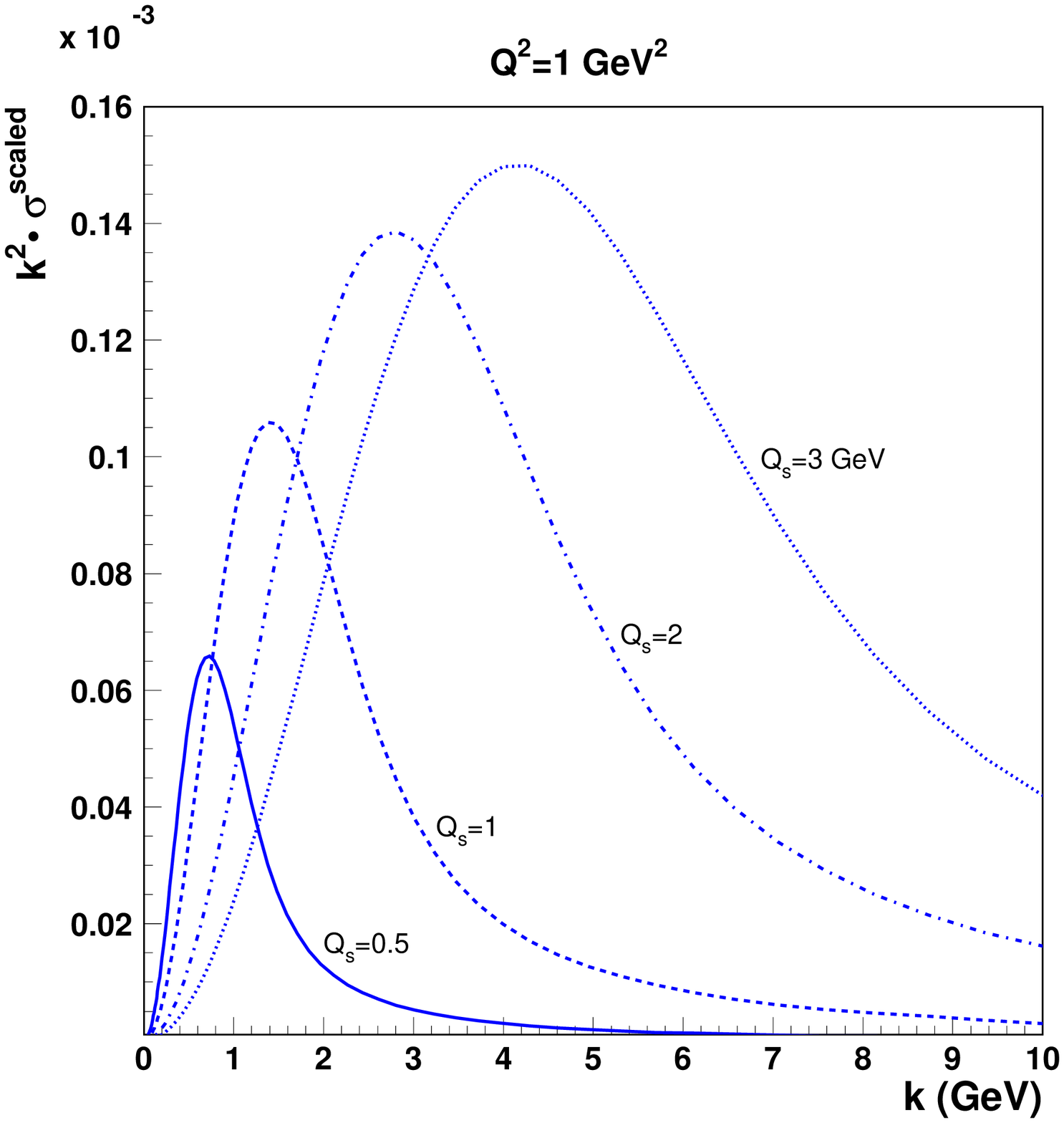,width=8.5cm}
\caption{The scaled cross section $\sigma^{scaled}$ and $k^2\sigma^{scaled}$ as 
a function of 
gluon transverse momentum $k$ for a fixed value of $Q^2=1~{\rm GeV}^2$ and four 
indicated values 
of the saturation scale $Q_s$.}
\end{center}
\label{F3}
\end{figure}

In Fig.3 we plot $\sigma^{scaled}$ and $k^2\sigma^{scaled}$
as a function of $k$, for fixed $Q^2\!=\!1~{\rm GeV}^2$ and four values of the 
saturation scale,
$Q_s=0.5, 1, 2, 3~{\rm GeV}$. As discussed in section \ref{sec:3}, independently 
of the form of the $S-$matrix,
$\sigma^{scaled}$ goes to a constant at small momenta while at large momenta 
$\sigma^{scaled}\sim 1/k^4$.
We check that this is the case on the first plot.
In the model with parton saturation (\ref{eq:S1}) the value of $\sigma^{scaled}$ 
as
$k\!\rightarrow\!0$ is strongly related to the saturation scale $Q_s$. This 
relation is better illustrated
on the second plot which represents the dimensionless quantity 
$k^2\sigma^{scaled}$. We see
that the transition region between two distinct behaviours at small and large 
$k^2$, which
features a marked bump, is linked to the value of $Q_s$.
It is interesting to explore
this observation hoping for the possibility to extract the saturation  scale 
from the measured dependence of
the diffractive cross section (\ref{obs}) on the gluon transverse momentum. Of 
course,
in the experimental situation  the gluon is seen as a jet. In the kinematic 
region of high diffractive mass
($\beta\ll 1$) the gluon jet is the closest to the edge of the rapidity gap. The 
contributions from quark-initiated jets close to the rapidity gap in such a 
kinematic domain are suppressed by $\log(1/\beta).$

\begin{figure}[ht]
\begin{center}
\epsfig{file=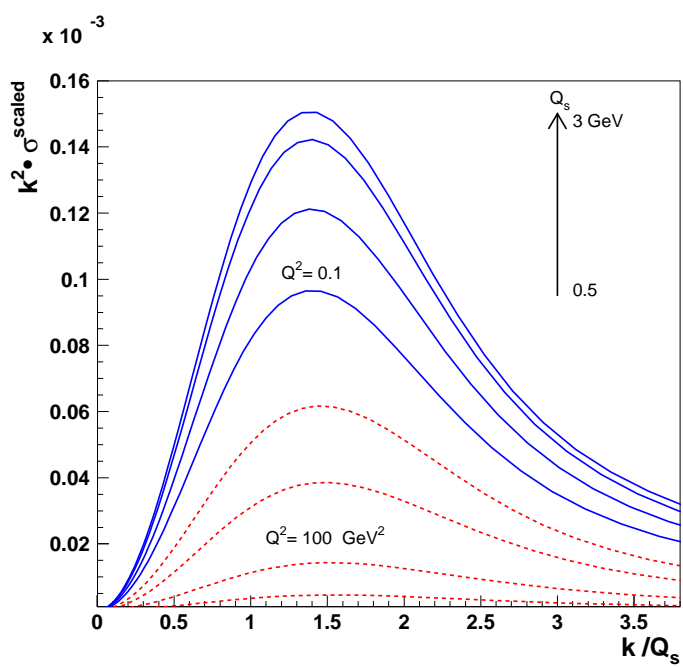,width=10cm}
\caption{The dimensionless cross section $k^2\sigma^{scaled}$ as a function of 
$k/Q_s$ for
two extreme values of $Q^2$ equal to $0.1$ and $100~{\rm GeV}^2$. The four 
curves for
each value of $Q^2$
correspond to four values of the saturation scale: $Q_s=0.5, 1, 2, 3~{\rm 
GeV}$.}
\end{center}
\label{F4}
\end{figure}

In order to quantify  the dependence of the position of the maximum of 
$k^2\sigma^{scaled}$ on
$Q_s$, we  plot  this cross section
as a function of the rescaled transverse momentum $k/Q_s$ for the four values of 
the saturation
scale indicated in Fig.~3.
In Fig.~4 we show the result of this study
for two extreme values of the photon virtuality, $Q^2=0.1$ and $100~{\rm 
GeV}^2$.
As clearly seen, the maximum for each curve is independent of $Q_s$ and $Q^2$ in 
a broad range of
considered values. From this figure we find that $k_{max}/Q_s=1.4\!-\!1.5$, thus 
within the saturation model, the maximum of $k^2\sigma^{scaled}(k)$ is 
proportional to the saturation scale $Q_s$ with a coefficient of proportionality 
independent of $Q^2$. In this way, if the saturation model is accurate, the 
diffractive gluon production in the domain of large diffractive mass offers 
a unique opportunity to determine the saturation scale $Q_s$ and its dependence 
on $\xp$.

As already discussed, in the experimental verification of the validity of our 
description in the $ep$ collisions at HERA, one should consider large-mass 
diffractive processes ($M_X\gg Q$) with a final-state configuration with a jet 
close to the rapidity gap: ${\rm X+jet+gap+p}$.
Then, the diffractive cross section (\ref{obs}) should be determined as a 
function of the jet tranverse momentum  for different values of $\xp$. Positions 
of the maximum of the measured cross section should be independent of $Q^2$, 
leading to the $\xp-$dependence of the saturation scale. The absolute value of 
the saturation scale depends of the coefficient of proportionality between 
$k_{max}$ and $Q_s$, which in our model equals $1.4-1.5$. Note that since 
$k_{max}$ is independent of $Q^2,$ a wide range of photon virtuality could be 
used to carry out this measurament, as long as one keeps $\beta\!\ll\!1.$

However, from the experimental point of view there exists an important 
limitation related to
the minimal value of the transverse momentum which could be measured for a jet.
In the most pesymistic scenario, considering even rather high values of the 
saturation scale, $Q_s(\xp)\sim\!1~{\rm GeV}$, it is unlikely
that the maximum $k_{max}$ of the cross section (\ref{obs}) can be seen at HERA.
Thus, to see the transition between the two different behaviours of the cross
section (\ref{obs}) seems like a major experimental challenge.

\begin{figure}[ht]
\begin{center}
\epsfig{file=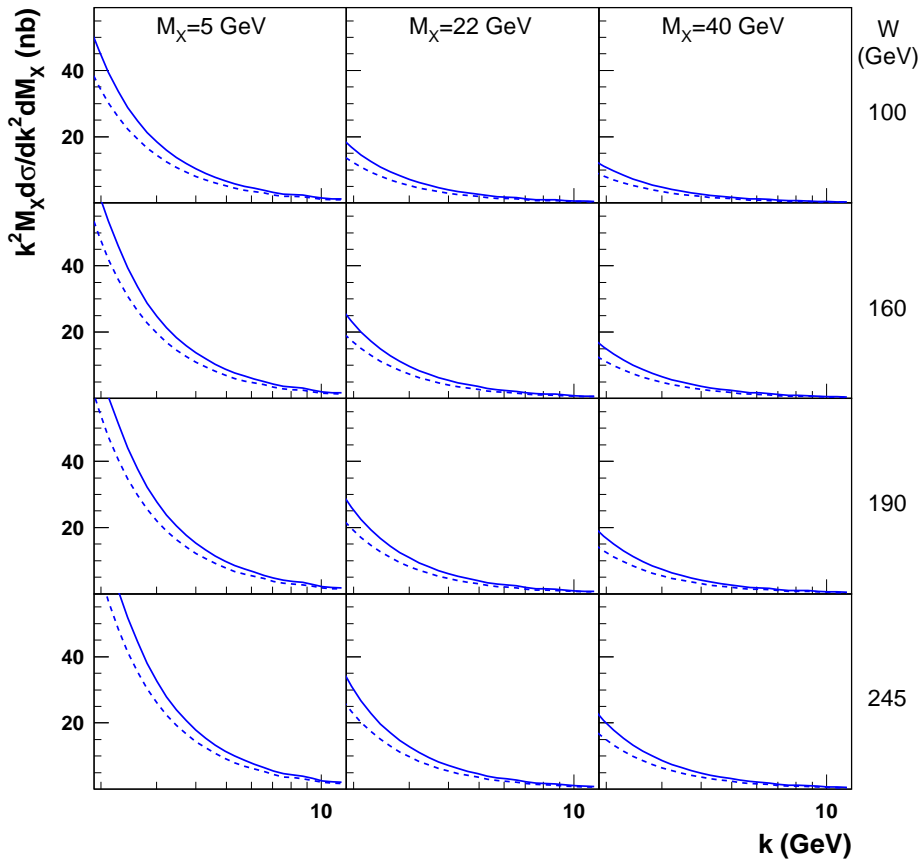,width=13 cm}
\caption{The cross-section $k^2M_Xd\sigma/dk^2dM_X$ as a function of the jet 
transverse 
momentum $k$ for $Q^2\!=\!0.1~{\rm GeV}$ and different values of diffractive 
mass $M_X$ and energy
$W$. Parameters
of saturation effects are from \cite{golec}: solid lines correspond to the case 
with light quarks
only, while the dashed  lines are predictions including the charm quark 
contribution.
}
\end{center}
\label{F5}
\end{figure}

In Fig.~5 we illustrated such a situation, when the saturation scale was taken 
from the model
\cite{golec} in which
\be\label{eq:golsat}
Q_s(\xp)=({x_0}/{\xp})^{\lambda/2}\ \mbox{GeV}
\ee
with the following parameters: $\lambda=0.288~(0.277)$ and $x_0=3.04\times 
10^{-4}~(0.4\times 10^{-4})$,
where the numbers in parenthesis refer to the
case where, in addition to the light quarks,
the charm quark is included in (\ref{phwf}). The cross-section (\ref{fincs}), 
shown in Fig.~5,
is computed for the above saturation scale and the parameters: 
$\alpha_s\!=\!0.15$ and
$\sigma_0=23.03~(29.12)~\mbox{mb}$ taken from \cite{golec}.
With these values and the saturation scale (\ref{eq:golsat}),
the diffractive cross-section (\ref{obs}) integrated over transverse momentum 
$k$,
$d\sigma^{diff}/dM_X$,  is well described \cite{munsho}.
The values of photon virtuality $Q^2$, energy $W$ and diffractive
mass $M_X$ indicated in Fig.~5 were taken form a recent analysis by the ZEUS 
Collaboration \cite{ZEUSrecent}.
One sees that, unfortunately, the data should always lie on the perturbative 
side of
the bump. However, it is not necessary to see the whole bump to confirm the 
influence of the
saturation scale on the results. In particular, there is a big difference in the 
rise towards the bump between the highest $\xp-$bin ($M_X\!=\!40~{\rm GeV}$ and 
$W\!=\!100~{\rm GeV}$) and the lowest $\xp-$bin ($M_X\!=\!5~{\rm GeV}$ and 
$W\!=\!245~{\rm GeV}$). A confirmation of such a behaviour would be a sign that 
the saturation region is indeed close and could lead to the determination of the 
saturation scale. If however this behaviour is not observed, it could reflect 
that our saturation model is incomplete, {\it e.g.} for example \eqref{eq:S2} 
neglects dipole correlations. It could also mean that in this process, unitarity 
does not come from saturation, but rather from soft physics~\cite{shaw}.

\section{Conclusions}
\label{sec:6}

Let us summarize the main results of this paper. We analysed the contribution of 
the $q\bar qg$ component of the virtual-photon wavefunction to the diffractive 
cross sections measured at HERA. In particular, we studied in detail
diffractive jet production in the large diffractive mass limit ($\beta\ll 1$) 
with a jet close to the rapidity gap. In such a case, the jet is initiated by 
the gluon.
We expressed the diffractive photon dissociation cross-section (\ref{obs}) in 
terms of dipole scattering matrices, formulae \eqref{eq:sigd}, (\ref{ampla}) and 
(\ref{cs}). We found that this cross section is strongly sensitive to unitarity 
constraints. In particular, independently of the form of the scattering 
matrices, the cross section (\ref{obs}) is a rising (falling) function of the 
final-state gluon transverse momentum in the limit $k\to 0$ ($k\to \infty$), 
with the maximum related to the scale at which unitarity effects become 
important.

In the context of saturation theory in which unitarity is realized 
perturbatively, the maximum is
determined by the saturation scale $Q_s$. Using the saturation parametrization 
\eqref{eq:S1} and \eqref{eq:S2} of
the scattering matrices, we verify that the relation between the maximum of 
(\ref{obs}) and the saturation scale is universal, i.e. independent of  $Q^2$. 
Therefore,
we propose the measurement of the diffractive jet production cross section
in $\gamma^*p$ collisions at HERA featuring the final-state configuration:
${\rm X+jet+gap+p}$ with a jet close to the rapidity gap. We argue that such a 
process offers an opportunity to extract the saturation scale from the 
experiment, provided a low enough jet transverse momentum can be measured.

\begin{acknowledgments}

We would like to thank St\'ephane Munier and Arif Shoshi for commenting on the 
manuscript.
C. M. wishes to thank the members of the department of theoretical physics at 
the INP in Krakow for their hospitality during his visit. This research has been 
supported by the grant from the Polish State Committee For Scientific Research, 
No.~1~P03B~028~28 and by the program ECONET No.~08155PC from the French 
Ministery of Foreign Affairs.

\end{acknowledgments}
\begin{appendix}

\section{Derivation of the amplitude $A$}

In this Appendix, we compute the amplitude (\ref{ampla}),
\be
A(k,x_0,x_1;x_{\pom})=\int\f{d^2z}{2\pi}\ e^{-ik.z}
\left[\f{z\!-\!x_0}{|z\!-\!x_0|^2}-\f{z\!-\!x_1}{|z\!-\!x_1|^2}\right]
\lr{S^{(2)}(x_0,z,x_1;x_{\pom})-S(x_0,x_1;x_{\pom})}\ ,
\ee
with the ${\cal S}-$matrices given by 
the saturation model (\ref{eq:S1}).
The virtual contribution is proportional to
\be
\int\f{d^2z}{2\pi}\ e^{-ik.z}
\left[\f{z\!-\!x_0}{|z\!-\!x_0|^2}-\f{z\!-\!x_1}{|z\!-\!x_1|^2}\right]=\,
\lr{{\rm e}^{-ik\cdot x_0}-{\rm e}^{-ik\cdot x_1}}\int\f{d^2z}{2\pi}\ 
{\rm e}^{-ik\cdot z}\f{z}{|z|^2}
\,=
-\frac{2k}{|k|^2}\,{\rm e}^{-ik\cdot b}
\sin\left(\half\, k\cdot x_{01}\right)\ .
\ee
One can then write
\be
A(k,x_0,x_1;x_{\pom})\,=\,\Theta(R_p-|b|)\lr{{\rm e}^{-ik\cdot x_0}\,I(k,x_{01})
\,-\,{\rm e}^{-ik\cdot x_1}\,I(k,-x_{01})\,+\,
\frac{2k}{|k|^2}\,{\rm e}^{-ik\cdot b}\,\sin\lr{\half\, {k\cdot x_{01}}}\,
{\rm e}^{-Q_s^2x_{01}^2/4}}
\label{ampl}
\ee
where we have introduced
\be
I(k,r)\,=\int\f{d^2z}{2\pi}\ {\rm e}^{-ik\cdot z}\,
\f{z}{|z|^2}\, {\rm e}^{-Q_s^2z^2/4}\, {\rm e}^{-Q_s^2(z+r)^2/4}\ .
\ee
Introducing $\theta$, the angle between $z$ and $k$, and $\phi$, the angle 
between
$r$ and $k,$ one has:
\be
I(k,r)\,=\,{\rm e}^{-Q_s^2r^2/4}\int\f{d|z|}{|z|}\,{\rm e}^{-Q_s^2z^2/2}\,
i\nabla_k\int\f{d\theta}{2\pi}\, 
{\rm e}^{-i|k||z|\cos\theta-\f12Q_s^2|z||r|\cos(\theta-\phi)}\ .
\ee
The angular integration gives:
\be
\int_0^{2\pi}\f{d\theta}{2\pi}\ 
{\rm e}^{-i|k||z|\cos\theta-\f12Q_s^2|z||r|\cos(\theta-\phi)}
\,=\,\ I_0\lr{|z|\sqrt{r^2Q_s^4/4+ik\cdot r\,Q_s^2\!-\!k^2}}
\ee
and differentiating with $i\nabla_k$, one obtains
\be
I(k,r)=-\f{ik+r\,Q_s^2/2}{\sqrt{r^2Q_s^4/4+ik\cdot r\,Q_s^2\!-\!k^2}}
\,\,{\rm e}^{-Q_s^2r^2/4}\int_0^\infty dz\,{\rm e}^{-Q_s^2z^2/2}\,
I_1\lr{z\sqrt{r^2Q_s^4/4+ik\cdot r\,Q_s^2\!-\!k^2}}\ .
\ee
Then performing the final integration, one gets
\be
I(k,r)=\f{ik+r\,Q_s^2/2}{k^2 - r^2\,Q_s^4/4 - ik\cdot r\,Q_s^2}
\ e^{-Q_s^2r^2/4}\lr{e^{-\!k^2/(2Q_s^2)+r^2 Q_s^2/8+ik\cdot r/2}-1}\ .
\label{inti}\ee
Inserting (\ref{inti}) into (\ref{ampl}) finally gives:
\bea\nonumber
\label{A9}
A(k,x_0,x_1;x_{\pom})\!&=&\!
\Theta(R_p-|b|)\,{\rm e}^{-ik\cdot b-Q_s^2x_{01}^2/4}
\bigg(\f{ik+x_{01}Q_s^2/2}{k^2-x_{01}^2 Q_s^4/4 - ik\cdot x_{01}\,Q_s^2}
\lr{{\rm e}^{- k^2/(2Q_s^2)+x_{01}^2Q_s^2/8}-e^{-ik\cdot x_{01}/2}}
\\\nonumber
\\
&-&\f{ik-x_{01}Q_s^2/2}{k^2 - x_{01}^2\,Q_s^4/4 + ik\cdot x_{01}Q_s^2}
\lr{{\rm e}^{-k^2/(2Q_s^2) + x_{01}^2Q_s^2/8}-{\rm e}^{ik\cdot x_{01}/2}}
+\frac{2k}{|k|^2}\sin\lr{\half{k\cdot x_{01}}}\bigg)\ .
\eea

\end{appendix}


\end{document}